# Observation of Non-Abelian Thouless Pump

Oubo You[1†], Shanjun Liang[2†], Biye Xie[1], Wenlong Gao[3], Weimin Ye[4], Jie Zhu[5,6*], Shuang Zhang[1,7*]

1. Department of Physics, University of Hong Kong, Hong Kong, China
2. Division of Science, Engineering and Health Studies, College of Professional and Continuing Education, Hong Kong Polytechnic University, Hong Kong, China
3. Department of Physics, Paderborn University, Warburger Straße 100, 33098 Paderborn, Germany
4. College of Optoelectronic Science and Engineering, National University of Defense Technology, Changsha, 410073, China
5. Institute of Acoustics, School of Physics Science and Engineering, Tongji University, Shanghai 200092, China
6. Department of Mechanical Engineering, Hong Kong Polytechnic University, Hong Kong, China
7. Department of Electronic & Electrical Engineering, University of Hong Kong, Hong Kong, China

Email: jiezhu@tongji.edu.cn; shuzhang@hku.hk

**Thouless pump is a one-dimensional dynamic topological effect that stems from the same topological mechanism as the renowned two-dimensional Chern insulators, with one momentum dimension replaced by a time variant evolution parameter. The underlying physics is abelian, concerning topologies of individual bands. By introducing multi-band non-abelian topology into pumping, much richer physics is expected that goes beyond the abelian counterpart. Here we report the observation of a non-abelian Thouless pump in an acoustic waveguide array with judiciously engineered coupling configurations. The observed non-abelian effect originates from time-evolution of eigen-states in $\mathbb{Z}_3$ pump cycles that traverse across several band degeneracies in the parameter space in a three-band system. Interestingly, the pump of the eigenstates exhibits non-commutativity, i.e. the final state depends critically on the order of the pumping paths. Our study paves ways for exploring and utilizing non-abelian dynamical effects in classical systems.**

Thouless pump provides robust ways to realize quantized transport of waves and particles, and it casts the static 2D quantum Hall effect onto 1D dynamic systems where one of the momentum dimensions is replaced by the evolution time or path parameter [1]. Thouless pump has been theoretically proposed [2-6] and experimentally realized [7-14] in various physical systems, such as electronics [2, 7], cold atoms [3, 8, 9], photonics [4, 10-12], acoustics [5, 13], and mechanics[14]. In the last few decades, as part of the booming development in the exploration of various topological phases of matter, the concept of Thouless pump has been extended to many different topological systems, such as higher-order topological pump [15, 16], disclination topological pump [17], charge pump in four-dimensional (4D) Hall effect [18, 19], $\mathbb{Z}_2$ spin pump [20], Floquet-Thouless Energy pump [21] and Fractional pump [22].

Non-Abelian physics is another important field has been extensively explored for decades in both high energy physics and condensed matter physics [23, 24]. It is manifested in various aspects such as non-abelian gauge field [25-27], non-abelian anyons and statistics [28, 29], non-abelian Berry curvature and Bloch oscillations [30-32], non-abelian nodal links and quaternion topological charges [33-36],

Young monopole and non-abelian Wilson line [37, 38] and linked Weyl surfaces [39, 40]. They are all rooted in the non-commutativity of operations which take the form of matrices (or tensors), with a simple example being two rotations in three-dimensional (3D) space about $x$ and $y$-axis. In particular, the non-abelian anyons are regarded as one of the promising ways for achieving fault-tolerant topological quantum computing [41].

In most of the previous studies, the evolution path of the topological pump was non-degenerate everywhere and hence the pumping operator between two different paths is always Abelian (i.e., they are commutative with each other). On the other hand, topological pump along paths traversing through band degeneracies has remained largely unexplored. Such pump is expected to go beyond the previous descriptions because the evolution operators are now expressed as matrices in which two pump paths can be potentially non-Abelian. In a very recent work [42], Thouless pump with doubly degenerate two-band system was theoretically studied, which features a U(2) gauge field with non-commutativity between different closed loops in the parameter space. However, no experimental observation of dynamic non-abelian Thouless pump effects has been reported thus far.

In this work, we report observation of the non-abelian Thouless pump in 1D acoustic waveguide array systems. The non-abelian property manifests itself through the transition among three interacting states [43] in the evolution path. When the order of evolution paths is interchanged, the topological pump evolves the same initial state into different final pumped states, hence exhibiting non-Abelian characteristics.

Here, the implementation of non-abelian Thouless pump is based on an array of acoustic waveguides with carefully tailored coupling configurations, as shown in Fig. 1a. Each unit cell has a cross section (Fig. 1b) featuring the coupling between three rectangular waveguides (labeled as $w_1$, $w_2$ and $w_3$) via interconnected narrow tubes. Here only the $p$ mode of the waveguide is considered, which has an antisymmetric field distribution along the long side of each waveguide. Hence, the offset of the interconnect tube from the center of the waveguide provides a means to control the sign and amplitude of the coupling coefficient between them. Besides the intra-cell coupling, inter-cell coupling also exists between the neighboring unit cells through the interconnecting tubes, as shown in Fig. 1c. Thus, considering both the intra-cell coupling and inter-cell coupling, the overall configuration is characterized by four normalized offset parameters, $S_1$, $S_2$, $S_3$ and $S_4$. The corresponding Hamiltonian of the system in terms of the offset parameters is provided in Supplementary Material Sec. II.

Fig. 1a shows the waveguide configurations corresponding to two different pumping loops, $C_3$ and $C'_3$, which are the basic operations that will be combined in different sequences to construct the non-abelian Thouless pump. The paths of $C_3$ and $C'_3$ in the four-dimensional $S_i$ parameter space are shown in Fig. 1d and 1e, respectively. The $C_3$ loop is confined to the $S_1S_3S_4$ subspace; it starts from the origin (0, 0, 0, 0), and passes sequentially through (0, 0, -1, 1), (1, 0, 0, 0), (0, 0, 1, 1) and finally returns back to the origin, and during the whole process, $S_2$ remains zero. On the other hand, the $C'_3$ loop is confined to the $S_2S_3S_4$ subspace, and the path starts from the origin, and then sequentially pass through (0, 0, 1, 1), (0, 1, 0, 0), (0, 0, -1, 1) and finally returns back to the origin, with $S_1$ remaining zero through the process. These two loops share the same base point located at the origin of the parameters space, where the three eigenstates are respectively the symmetric and antisymmetric modes formed by the p modes of the top and bottom ($w_1$ and $w_2$) waveguides (labelled as $|+\rangle$ and $|-\rangle$ respectively), and the mode that is only localized in the linking waveguide $w_3$ ($|l\rangle$), as shown in Fig. 1f.

Because the propagation distance $z$ is treated as time, the propagation constant $k_z$ functions effectively as the energy in 1D system. The operation of $C_3$ loop to the initial states can be illustrated by the band evolution along the $C_3$ loop, as shown in Fig. 2a. The transition between different states is enabled by the band crossings located at the origin and at (1, 0, 0, 0) in the parameter space. Starting from the origin in the parameter space, the three initial states $|+\rangle$ (enclosed by a blue square), $|-\rangle$ (enclosed by a green square), and $|l\rangle$ (enclosed by a red square) will evolve into $|l\rangle$, $|+\rangle$ and $|-\rangle$ respectively, after a single $C_3$ loop. Hence the three states form a permutation cycle under $C_3$ operation, as illustrated by Fig. 2b. Note that all the transition occurs within the same unit cell, because $S_2$ remains zero along the $C_3$ loop, i.e., there is no coupling between neighboring unit cells during the evolution. Therefore, the projection of the operation on the evolved three-band subspace can be mathematically expressed as

$$C_3 = \sum_N (|l,N\rangle\langle +,N| + |-,N\rangle\langle l,N| + |+,N\rangle\langle -,N|), \tag{1}$$

with N denotes the site number. Based on Eq. (1), the original states will return back to themselves after three consecutive $C_3$ loops, i.e. the evolution loop $C_3$ represents a $\mathbb{Z}_3$ pump process. Similarly, the $C'_3$ loop provides another permutation cycle between the three states but of opposite direction as $C_3$, as illustrated by Fig. 2c. Importantly, $C'_3$ operation involves the permutation among states $|+\rangle$ and $|-\rangle$ of one unit cell and $|l\rangle$ state in the unit cell to the left, and it can be mathematically expressed as,

$$C'_3 = \sum_N (|-,N\rangle\langle +,N| + |l,N-1\rangle\langle -,N| + |+,N\rangle\langle l,N-1|). \tag{2}$$

Eqs. (1) and (2) show that each single evolution is equivalent to a $\mathbb{Z}_3$ pump, and there is no lattice shift if the evolution path only consists of multiples of solely $C_3$ or $C'_3$ because the $\mathbb{Z}_3$ topology leads to $C_3^3 = C'^3_3 = I$. However, if one combines these two different $\mathbb{Z}_3$ pump loops together, the lattice index in the expression can vary and the states can be pumped to the left or the right, or trapped in any arbitrary manner. A unified description of the effects of these two pump loops is summarized in the map shown in Fig. 2d. Based on this map, one can always construct a path to achieve pump from a state in a particular unit cell to any states in a different unit cell.

To elucidate such non-abelian pump, we consider two synthetic $\mathbb{Z}_3$ pump paths, each consisting of three $C_3$ loops and one $C'_3$ loop but arranged in different sequences, as indicated by the black solid and dotted lines in Fig. 2d. For a given initial state $|+\rangle$, following the labelled sequence from 1 to 4, it will evolve into the final state of $|-\rangle$ for both paths, but the final state is pumped to the right for the path with sequence of $C_3 C_3 C'_3 C_3$ (**Path1**, corresponding to the dashed line in Fig. 2d) and trapped for the other path with a sequence of $C'_3 C_3 C_3 C_3$ (**Path2**, corresponding to the solid line). The simulation results on waveguide configurations corresponding to these two evolution paths are provided in Supplementary Material Sec. III, confirming the pump and trapping behaviors for **Path1** and **Path2**, respectively.

As shown in Fig. 1d and 1e, $C_3$ and $C'_3$ each contains four line segments, and each line segment has a length of P/4 with P denoting the total length of $C_3/C'_3$. The overall propagation length of the system for each path, consisting of 16 segments, would be very long, which hampers the experimental observations due to the significant intrinsic attenuation after a long path. In order to facilitate the experimental implementation, we employ a cut-and-join process, in which different locations of the waveguides where the field distributions are identical to each other are joined together, and the evolution path in-between is omitted, as detailed in Supplementary Material Sec. III. By doing this, the length of the waveguide array is significantly reduced without affecting the pumping results. The final shortened configurations of the waveguide array each consists of two components $L_1$ and $L_2$, which, when joined together in different orders, leads to different pumping behaviors. The corresponding acoustic pressure distributions simulated at an excitation frequency of 9200Hz for both configurations are shown in Fig. 3a. It is shown that, despite the greatly shortened propagation length, the two pumping operations formed by the combination of $L_1$ and $L_2$ clearly exhibit the same pump behaviors as the original paths consisting of four $C_3/C'_3$ loops - the state is shifted to the right for one configuration ($L_2L_1$), and trapped for the other configuration with the order of the two operations swapped ($L_1L_2$), hence a clear evidence of the non-abelian pumping behavior.

In the experiment, the acoustic waveguide is 3D-printed structure surrounded by 5mm thick wall consisting of material Somos EvoLVe 128, which can be treated as rigid boundary for the airborne sound wave traveling inside. At the input ports, a pair of out-of-phase tiny-size armature drivers, excite the $p$ mode acoustic wave inside the rectangular waveguide. At the ends of the waveguide, porous absorptive cotton is used as the non-reflection boundary. In order to measure the evolution of the field along the propagation direction, an array of tiny ladder holes is drilled into the top and bottom sides of each waveguide. The holes are sealed by 3D printed plugs to ensure air tightness and is only unplugged when being measured. Experimental results of the acoustic pressure distribution at the top of both samples are shown in Fig. 3b for the two configurations corresponding to $L_2L_1$ and $L_1L_2$, respectively. The measured results clearly show the expected non-abelian pumping behaviors, which are in good agreement with the simulation results. More details of experimental setup and data are given in Supplementary Material Sec. IV, V and VI.

In summary, we have experimentally demonstrated non-abelian Thouless pump in an acoustic waveguide system. Unlike previous proposals in which the state evolution takes place in perfectly degenerate bands driven by non-abelian Berry connections, the non-abelian pump demonstrated here arises from combination of different $\mathbb{Z}_3$ pump cycles enabled by band crossing in three connected bands. Hence, the Wilczek-Zee phase experiences an abrupt change at the band crossings, instead of being continuously accumulated along the path. Our study unequivocally demonstrates the non-Abelian characteristics of Thouless pump and paves ways for exploration of new physical phenomena and device applications based on non-abelian dynamical effects in classical systems.

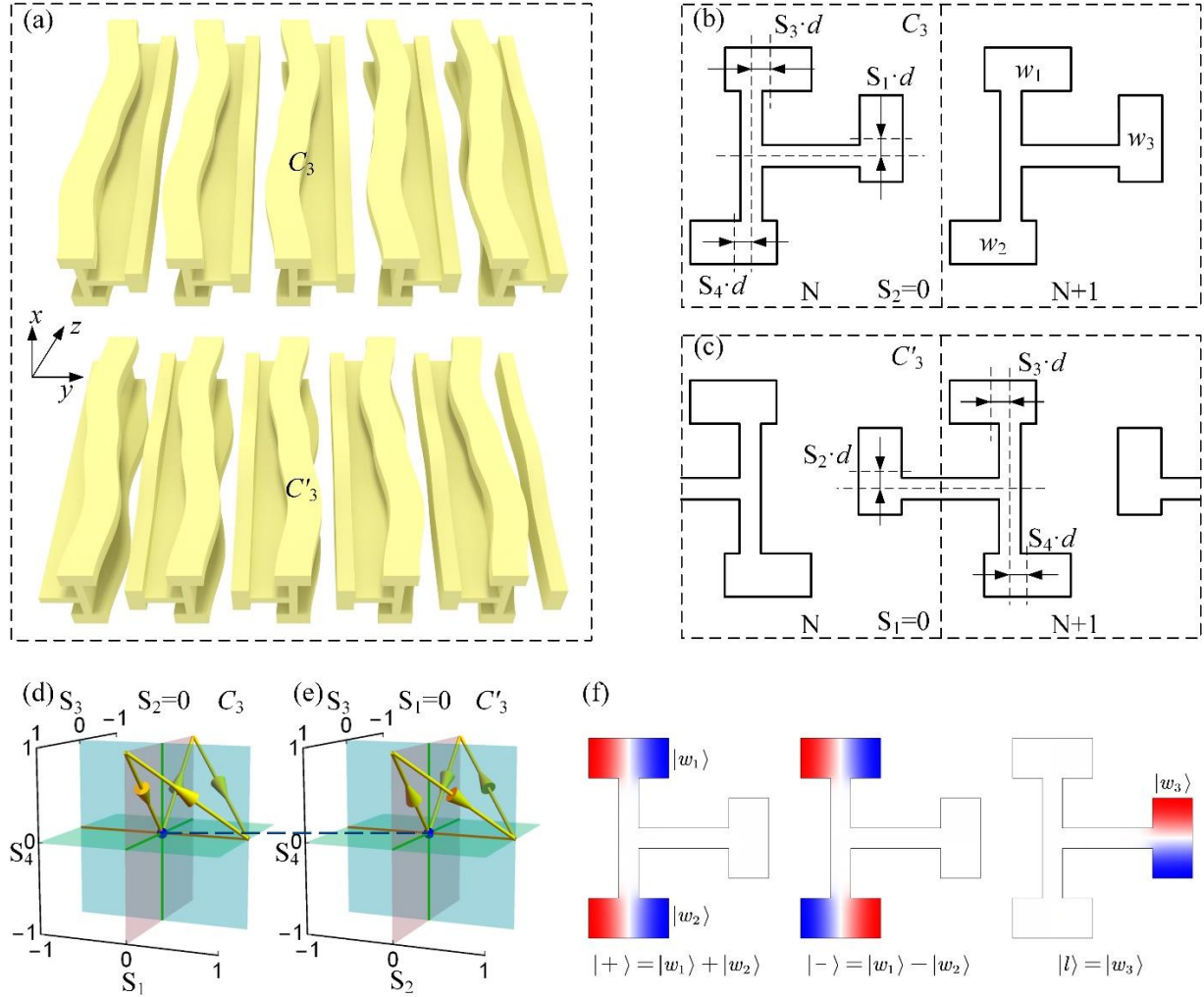

Fig. 1 Configurations of the acoustic waveguide arrays for the study of pumping. (a). Configurations of 1D periodic acoustic waveguide array for pumping loop $C_3$ and $C'_3$. (b, c). Illustration of the geometric parameters of the cross section of the coupled waveguide array. The parameters $S_i$'s are the offsets between the centers of the rectangular waveguides and T-shaped tube. (d). Illustration of the pumping loop (indicated by yellow arrows) in the parameter space, which resides in the $S_1S_3S_4$ subspace ($S_2=0$). (e). Illustration of the pumping loop in the parameter space, which resides in the $S_2S_3S_4$ subspace ($S_1=0$). The origins in (d) and (e) represent the same point in the full parameter space $S_1S_2S_3S_4$ which is also the origin. (f). Field distribution for mode $|+\rangle, |-\rangle$ and $|l\rangle$.

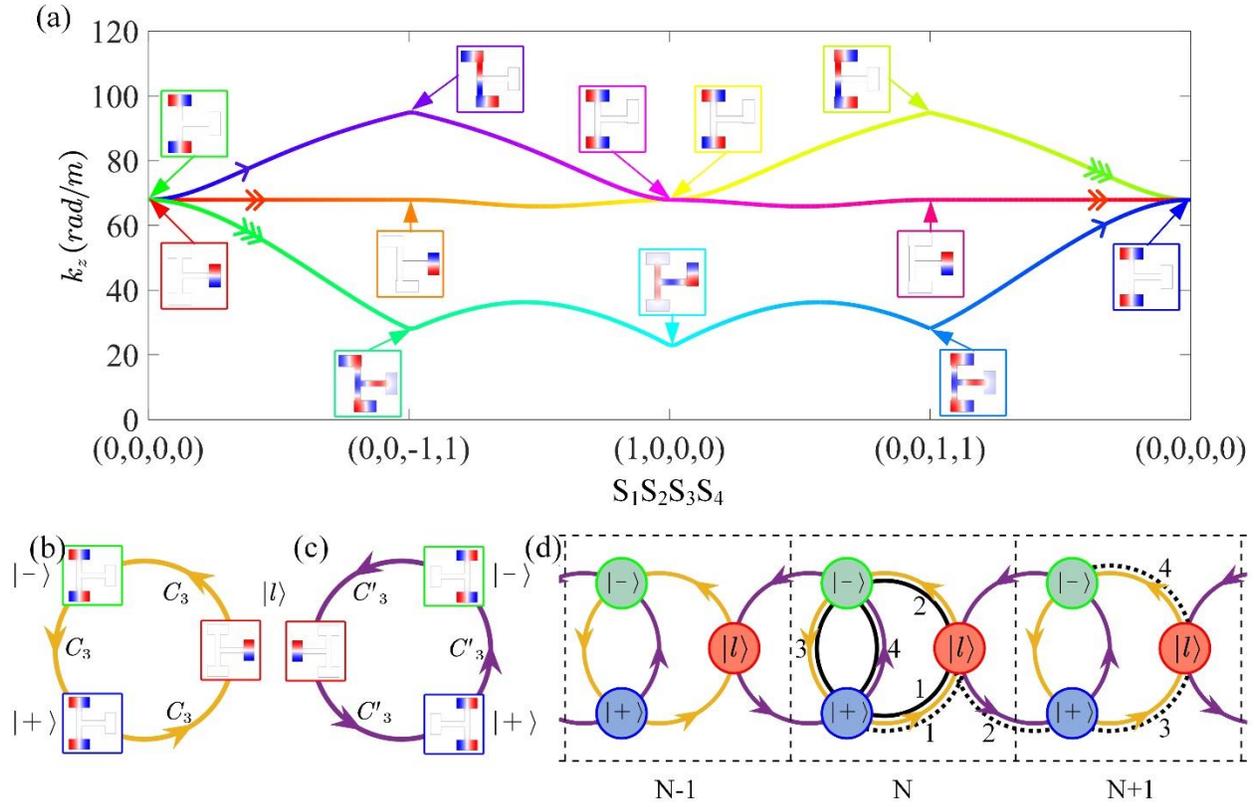

Fig. 2 Illustration of the $Z_3$ pumping cycles and non-Abelian Thouless pump. (a) The band structure along $C_3$ loop at frequency 9500Hz. The evolution directions are indicated by the arrows, and the corresponding eigenstates are provided along the evolution path. (b, c) A concise description of the operation of $C_3$ and $C'_3$. (d) The combination of $C_3$ and $C'_3$ on 1D periodic waveguides. Two paths $C_3C_3C'_3C_3$ and $C'_3C_3C_3C_3$ are indicated by dotted and solid black lines.

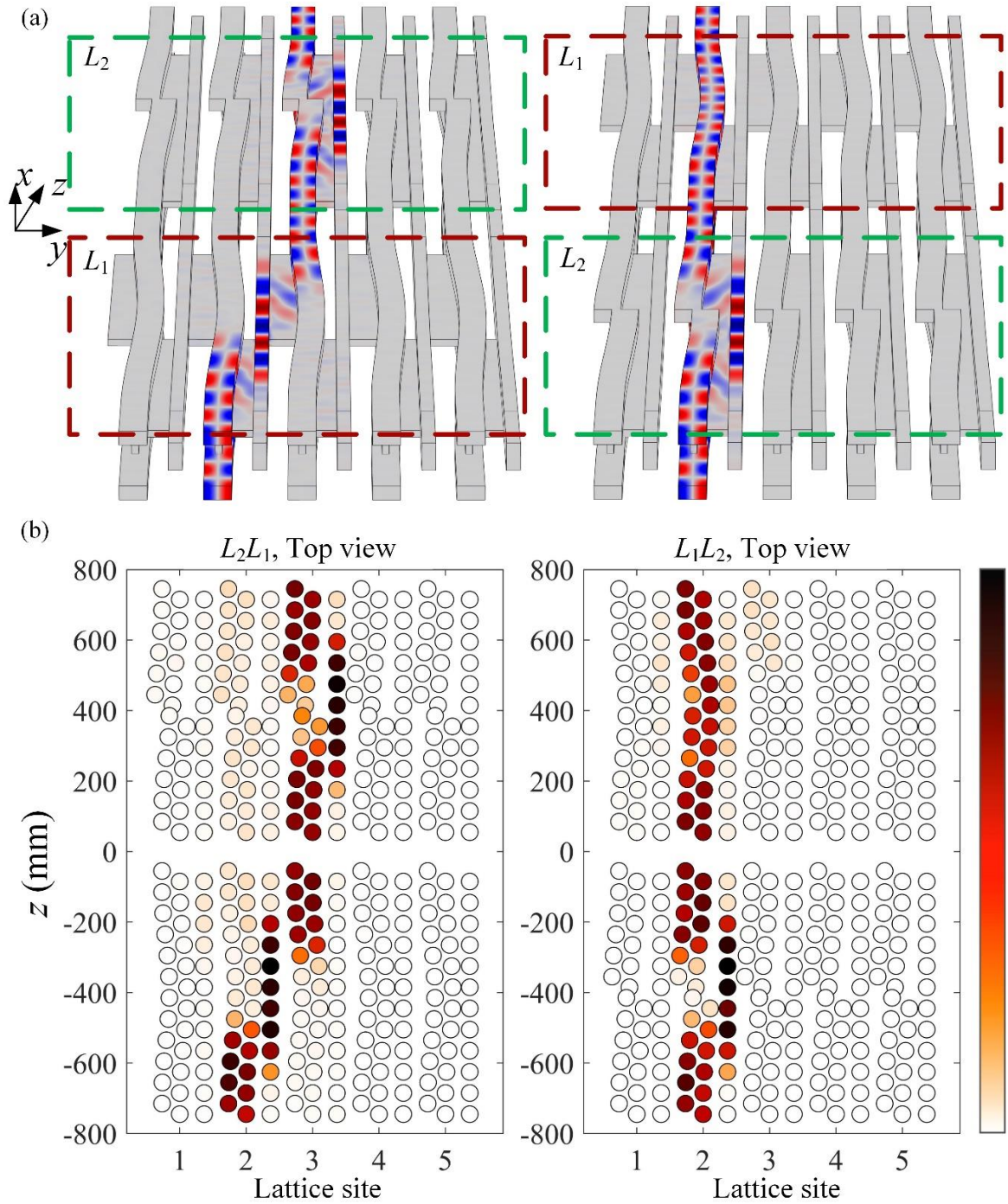

Fig. 3 Full wave simulation and measurement of the Thouless pumping. (a) The simulated acoustic pressure distribution along the paths $L_2L_1$ (left) and $L_1L_2$ (right), which are equivalent to the full evolution paths $C_3C_3C'_3C_3$ and $C'_3C_3C_3C_3$, respectively. The simulation clearly shows that the final state is pumped to the right for the paths $L_2L_1$, and trapped for $L_1L_2$. (b) The measured pressure at discrete locations along the path $L_2L_1$ and $L_1L_2$, which agree very well with the simulation results.